\documentclass[a4paper,12pt]{article}
\usepackage{graphicx,epsfig,geometry}
\newcommand{\ra}{\rightarrow}
\newcommand{\lam}{\Lambda^0}

\newcommand{\sig}{\Sigma^+}
\newcommand{\si}{\Sigma^0}

\begin{document}

\epsfysize3cm
\epsfbox{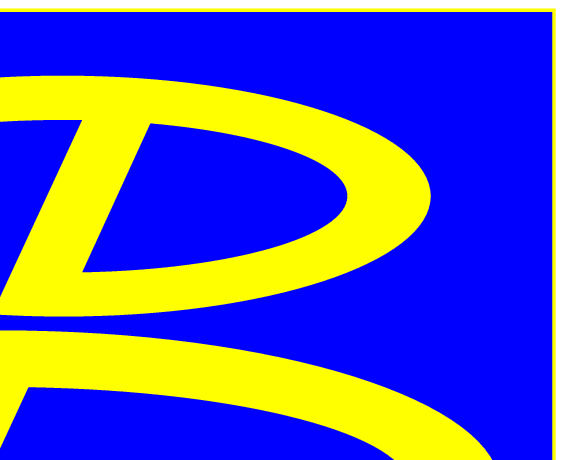}    
\vskip -3cm
\noindent
\hspace*{4.5in}BELLE Preprint 2001-17 \\
\hspace*{4.5in}KEK Preprint 2001-137 \\


\vspace*{3cm}

\begin{center}
{\Large \bf 
Observation of Cabibbo-suppressed
and W-exchange\\[10pt]
 {\boldmath $\Lambda_{c}^+$} baryon decays\footnote{
submitted to Phys. Lett. B.}
}
\end{center}

\vspace*{1cm}

\begin{center}
{\large The Belle Collaboration}\\

\vspace*{1cm}

  K.~Abe$^{9}$,               
  K.~Abe$^{42}$,              
  R.~Abe$^{32}$,              
  T.~Abe$^{43}$,              
  I.~Adachi$^{9}$,            
  Byoung~Sup~Ahn$^{17}$,      
  H.~Aihara$^{44}$,           
  M.~Akatsu$^{25}$,           
  Y.~Asano$^{49}$,            
  T.~Aso$^{48}$
% ...
\end{center}

\small
\begin{center}
% ...
$^{9}${High Energy Accelerator Research Organization (KEK), Tsukuba}\\
% ...
$^{17}${Korea University, Seoul}\\
% ...
$^{25}${Nagoya University, Nagoya}\\
% ...
$^{32}${Niigata University, Niigata}\\
% ...
$^{42}${Tohoku Gakuin University, Tagajo}\\
$^{43}${Tohoku University, Sendai}\\
$^{44}${University of Tokyo, Tokyo}\\
% ...
$^{48}${Toyama National College of Maritime Technology, Toyama}\\
$^{49}${University of Tsukuba, Tsukuba}\\
% ...
\end{center}

\normalsize

\normalsize



\newpage

\begin{abstract}

We present measurements of the Cabibbo-suppressed decays  
$\Lambda_{\rm c}^+\ra \lam K^+$ and $\Lambda_{c}^+\ra \si K^+$
(both first observations), $\Lambda_{c}^+\ra \Sigma^+ K^+\pi^-$ 
(seen with large statistics for the first time), 
$\Lambda_{c}^+ \ra p K^+ K^-$ and $\Lambda_{c}^+\ra p \phi$
(measured with improved accuracy).
Improved branching ratio measurements for the decays
$\Lambda_{c}^+ \ra \sig K^+ K^-$ and $\Lambda_{c}^+\ra \sig \phi$,
which are attributed to W-exchange diagrams, are shown. 
We also present the first evidence for $\Lambda_{c}^+ \ra \Xi(1690)^0 K^+$ 
and set an upper limit on the non-resonant decay $\Lambda_{c}^+ \ra \sig K^+ K^-$.
This analysis was performed using 32.6~fb$^{-1}$ of data collected 
by the Belle detector at the asymmetric $e^+ e^-$ collider KEKB. 

\end{abstract}


\section{Introduction}

Decays of charmed baryons, unlike charmed mesons, 
are not colour or helicity suppressed, allowing us to investigate the
contribution of W-exchange diagrams. 
There are also possible interference effects due to the presence of identical
quarks. This makes the study of these decays a useful tool to test 
theoretical models that predict exclusive decay rates~\cite{theory}.

During the past several years there has been significant progress in the
experimental study of hadronic decays of charmed baryons.
New results on masses, widths, lifetimes and decay asymmetry 
parameters have been published by various experiments~\cite{PDG}.
However the accuracy of branching ratio measurements does not exceed 30\% for
many Cabibbo-favoured modes: for Cabibbo-suppressed and W-exchange dominated
decays, the experimental accuracy is even worse.
As a result, we are not yet able to conclusively distinguish between the decay
rate predictions made by different theoretical models.

In this paper we present a study of $\Lambda_{c}^+$ baryons produced in the
$e^+ e^- \ra q {\bar q}$ continuum at Belle, relying on the excellent particle
identification capability of the detector to measure decays with 
kaons in the final state.
We report the first observation of the Cabibbo-suppressed decays
$\Lambda_{c}^+ \ra \lam K^+$ and $\Lambda_{c}^+ \ra \si K^+$, and the
first observation of $\Lambda_{c}^+ \ra \sig K^+ \pi^-$ with large statistics.
(Here and throughout this paper, the inclusion of charge-conjugate states is
implied.)
We present improved measurements of the Cabibbo-suppressed decays
$\Lambda_{c}^+ \ra p K^+ K^-$ and $\Lambda_{c}^+\ra p \phi$,
and the W-exchange decays  $\Lambda_{c}^+ \ra \sig K^+ K^-$ and  
$\Lambda_{c}^+\ra \sig \phi$;
we also report the first evidence for $\Lambda_{c}^+ \ra \Xi(1690)^0 K^+$,
and set an upper limit on non-resonant $\Lambda_{c}^+ \ra \sig K^+ K^-$ decay.


\section{Data and Selection Criteria}
\label{section-selection}

The data used for this analysis were taken on the $\Upsilon(4S)$ 
resonance and in the nearby continuum
using the Belle detector at the asymmetric $e^+ e^-$ collider KEKB. 
The integrated luminosity of the data sample is equal to 32.6~fb$^{-1}$. 

Belle is a general purpose detector based on a 1.5 T superconducting solenoid;
a detailed description can be found elsewhere~\cite{BELLE_DETECTOR}.
Tracking is performed with a silicon vertex detector (SVD) composed of three
concentric layers of double-sided silicon strip detectors, and a 50 layer
drift chamber.
Particle identification for charged hadrons, important for the measurement of
final states with kaons and/or protons,  is based on the combination of  
energy loss measurements $(dE/dx)$ in the drift chamber,
time of flight measurements and aerogel {\v C}erenkov counter information. 
For each charged track, measurements from these three subdetectors are 
combined to form $K/\pi$ and $p/K$ likelihood ratios in the range from 0 to 1,
$${\rm P}(K/\pi) = \mathcal{L}(K)/(\mathcal{L}(K) + \mathcal{L}(\pi)),~~ 
{\rm P}(p/K) = \mathcal{L}(p)/(\mathcal{L}(p) + \mathcal{L}(K)),$$
where $\mathcal{L}(p)$, $\mathcal{L}(K)$ and $\mathcal{L}(\pi)$ are the likelihood values
assigned to each identification hypothesis for a given track.

For the analyses presented here, we require P($K/\pi) < 0.9$ for pions, 
P($K/\pi) > 0.6$ for kaons, and P($p/K) > 0.9$ for protons, unless
stated otherwise.
Candidate $\pi^0$'s are reconstructed from pairs of photons detected in the 
CsI calorimeter, with a minimum energy of 50~MeV per photon. The interaction
point (IP) coordinates in the $r-\phi$ plane are determined from beam profile 
measurements. Other particles are identified as follows:
\begin{itemize}
  \item	$\lam$ are reconstructed in the decay mode $\lam \rightarrow p \pi^-$,
	fitting the $p$ and $\pi^-$ tracks to a common vertex and  
	requiring an invariant mass in a $\pm 3$~MeV$/c^2$ ($\approx \pm 3\sigma$)
	interval around the nominal value.
	The likelihood ratio cut on the proton is relaxed to P$(p/K) > 0.4$.
	We then make the following cuts on the $\lam$ decay vertex: 
	\begin{itemize}
	  \item	the closest distance of approach along the beam direction
                between the proton and pion tracks must be less than 1~cm;
	  \item	the distance between the decay vertex and the interaction point
		in the $r-\phi$ plane must be greater than 1~mm;
	  \item	the cosine of the angle in the $r-\phi$ plane
                between the $\lam$ momentum vector and the
		vector pointing from the IP to the decay vertex
		must be greater than 0.995.
	\end{itemize}
  \item	$K^0_{\rm S}$ are reconstructed in the decay mode $K^0_{\rm S} \rightarrow \pi^+ \pi^-$,
	fitting the $\pi^+$ and $\pi^-$ tracks to a common vertex and  
	requiring an invariant mass in a $\pm 7$~MeV$/c^2$ ($\approx \pm 3\sigma$)
	interval around the nominal value.
	We then make the same vertex cuts as in the $\lam$ case. 
  \item	$\sig$ are reconstructed in the decay mode $\sig \rightarrow p \pi^0$,
	requiring an invariant mass within $\pm 10\, \mathrm{MeV}/c^2$
	($\approx \pm 2 \sigma$) of the nominal value.
        We require the proton to have at least one hit in the SVD, to improve its
        impact parameter resolution with respect to the IP; 
	we then require the impact parameter in the $r-\phi$ plane
	to be greater than $200\,\mu\mathrm{m}$,
        to make sure the $\sig$ vertex is displaced from the IP.
  \item	$\si \rightarrow \lam \gamma$ decays are formed using identified $\lam$
	and photons with calorimeter cluster energies above $0.1$ GeV;
	we accept candidates with invariant masses within $\pm 6$~MeV$/c^2$
	($\approx \pm 1.5\sigma$) of the nominal value.
\end{itemize}

To suppress combinatorial and $B\overline{B}$ backgrounds,
we require $\Lambda_{c}^+$ candidates to have scaled momentum
$x_p=p^\ast/p^\ast_{\rm max} > 0.5$; here 
$p^\ast$ is the reconstructed momentum of the $\Lambda_c^+$ candidate
in the $e^+e^-$ center of mass, 
and
$p^\ast_{\rm max} = \sqrt{s/4-M^2}$,
where $\sqrt{s}$ is the total center of mass energy
and $M$ is the
reconstructed mass of the $\Lambda_{c}^+$ candidate.
In modes where there are two or more charged tracks at the $\Lambda_{c}^+$
vertex, we perform a vertex fit and require $\chi^2/\mathrm{n.d.f.} < 9$.

In the various mass fits described below, the central value and width of
the signal peaks are always allowed to float, unless stated otherwise. 
Wherever the final state includes a hyperon, we improve the invariant mass resolution
by plotting the corrected mass difference, \emph{e.g.}\ 
$M(\sig K^+ K^-) - M(\sig) + M_{\sig}^{\rm PDG}$
instead of $M(\sig K^+ K^-)$.


\section{Observation of the decays
	{\boldmath $\Lambda_{\lowercase {c}}^+ \rightarrow \lam K^+$} and
	{\boldmath $\Lambda_{\lowercase {c}}^+\ra \Sigma^0 K^+$}}

\begin{figure}[h]
\centering
\begin{picture}(550,200)
\put(80,40){\rotatebox{90}{\large \bf Entries/(5 MeV/{\boldmath $c^2$})}}
\put(145,0){\boldmath $M(\lam K^+)-M(\lam)+1.116,~~~{\rm [GeV}/c^2]$} 
\put(110,10){\includegraphics[width=0.6\textwidth]{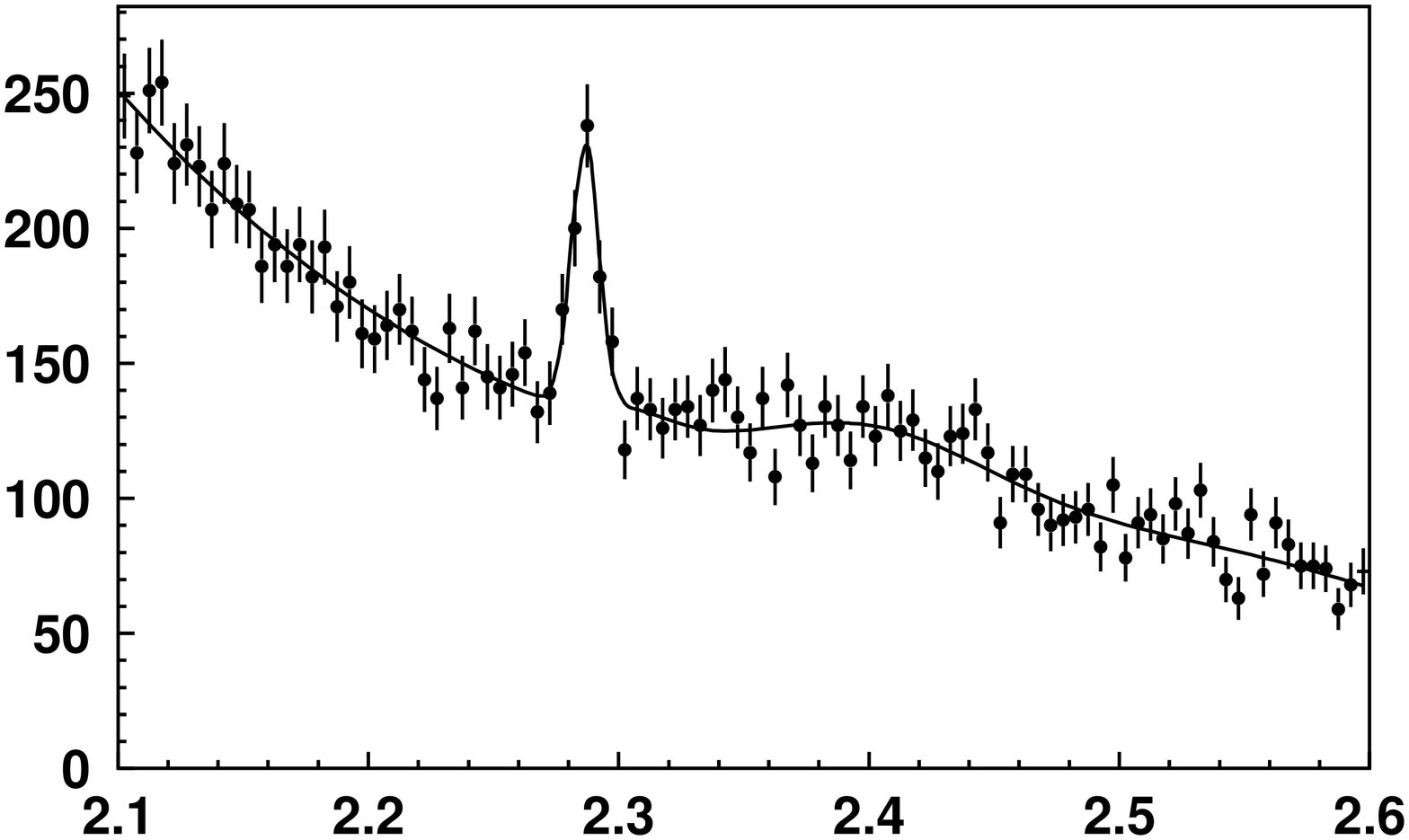}}
\end{picture}
\caption{$\Lambda_{\lowercase {c}}^+\ra \lam K^+$: invariant mass spectrum of
	the selected $\lam K^+$ combinations.
The broad structure to the right of the signal peak, due to 
$\Lambda_c^+ \ra \lam \pi^+$ and $\Lambda_c^+ \ra \si \pi^+$ decays, is included
in the fit.
}
\label{fig_lamc_lam0k_1}
\end{figure}

The Cabibbo-suppressed decay $\Lambda_{\lowercase {c}}^+\ra \lam K^+$ has not
been previously observed. Reconstructing $\lam K^+$ combinations as described
in Section~\ref{section-selection}, we see a clear signal at the $\Lambda_{c}^+$
mass, as shown in Fig.~\ref{fig_lamc_lam0k_1}.

To study backgrounds due to Cabibbo-allowed decays, we select a second
sample with a reversed identification requirement P$(K/\pi)<0.1$ applied to the
``kaon''. In the mass spectrum of this sample, where the kaon mass hypothesis is still
used, we see a broad structure centered around $2.4\, \mathrm{GeV}/c^2$,
produced by
$\Lambda_{c}^+\ra \lam \pi^+$ and $\Lambda_{c}^+\ra \Sigma^0 \pi^+$ decays. 
We fit this distribution
using two Gaussians 
(to model the $\Lambda_{c}^+\ra \lam \pi^+$ and $\Lambda_{c}^+\ra \Sigma^0 \pi^+$
contributions), and a second order polynomial (to model the broad reflections
and the remaining background).
The shape of this function is then used to model the $\Lambda_c^+ \ra \lam \pi^+ (\si \pi^+)$
background in the main sample.
The remaining combinatorial background in Fig.~\ref{fig_lamc_lam0k_1} is 
represented using a second order polynomial, and the 
$\Lambda_{c}^+\ra \lam K^+$ signal is described by a Gaussian 
with width $\sigma=5.4\, \mathrm{MeV}/c^2$ (fixed from Monte Carlo);
the result of the fit is shown by the superimposed curve.
We find a yield of $265 \pm 35$ $\Lambda_{c}^+\ra \lam K^+$ decays,
the first observation of this decay mode. 

For normalization, we use the decay $\Lambda_{c}^+\ra \lam\pi^+$.
The $\Lambda_{c}^+\ra \lam\pi^+$ mass
distribution is 
fitted with a Gaussian for the signal and 
a second order polynomial for the background.
We find $4550 \pm 111$ events. 
The relative reconstruction efficiency was determined using, Monte Carlo simulation (MC),
to be
$\epsilon(\Lambda_{c}^+\ra \lam K^+)/\epsilon(\Lambda_{c}^+\ra \lam\pi^+) 
 = 0.79$. Using this value, we extract 
$$
   \frac{\mathcal{B}(\Lambda_{c}^+\ra \lam K^+)}
  {\mathcal{B}(\Lambda_{c}^+\ra \lam\pi^+)} = 0.074\pm 0.010 \pm 0.012;
$$ 
the first error is statistical, and the second is systematic. We provide a detailed description of the 
sources of systematic error for this and other measured decay modes in Section~\ref{errors}.

\begin{figure}[h]
\centering
\begin{picture}(550,200)
\put(80,40){\rotatebox{90}{\large \bf Entries/(5 MeV/{\boldmath $c^2$})}}
\put(145,0){\boldmath $M(\Sigma^0 K^+)-M(\Sigma^0)+1.192,~~~[{\rm GeV}/c^2]$} 
\put(110,10){\includegraphics[width=0.6\textwidth]{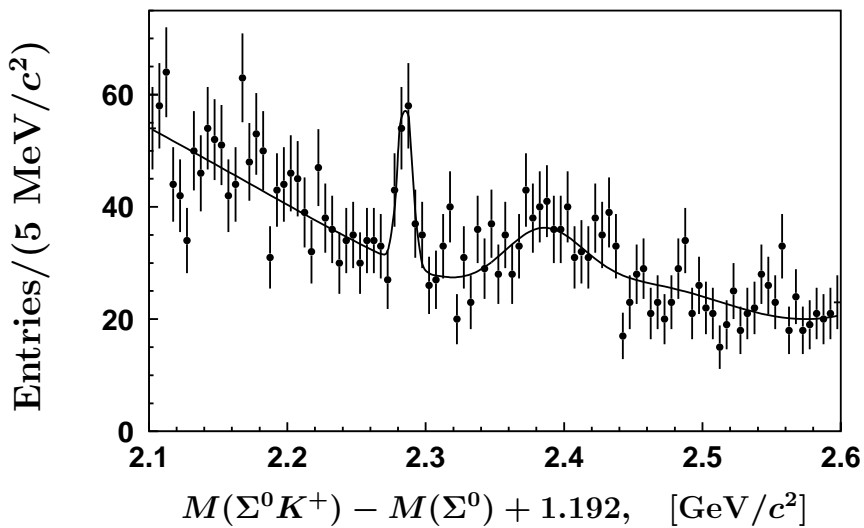}}
\end{picture}
\caption{$\Lambda_{c}^+ \ra \Sigma^0 K^+$: invariant mass spectrum of the
	selected $\Sigma^0 K^+$ combinations.
The broad structure to the right of the signal peak, due to 
$\Lambda_c^+ \ra \lam \pi^+$ and $\Lambda_c^+ \ra \si \pi^+$ decays, is included
in the fit.
}
\label{fig_lamc_sig0k_1}
\end{figure}

The Cabibbo-suppressed decay $\Lambda_{c}^+\ra \Sigma^0 K^+$ is reconstructed in
a similar way, with the scaled momentum cut tightened to $x_p>0.6$ to suppress
the large background due to soft photons. The invariant mass distribution of
the selected $\si K^+$ candidates is shown in Fig.~\ref{fig_lamc_sig0k_1}:
a peak is seen at the $\Lambda_{c}^+$ mass, and a reflection due to misidentified
two-body Cabibbo-allowed $\Lambda_{c}^+$ decays is seen at higher masses. 
The superimposed curve shows the result of a fit following the 
method described for $\lam K^+$, with the exception that in this case the width of the
signal Gaussian is fixed from the MC to 
$\sigma=5.0$~MeV$/c^2$.
We find $75\pm 18$ $\Lambda_{c}^+\ra \si K^+$ events, the first observation of
this decay mode. 
For normalization, we use the decay $\Lambda_{c}^+\ra \si \pi^+$.
We fit the distribution with a Gaussian for the signal,
a second Gaussian to describe the broad enhancement due to $\Lambda_{c}^+\ra \lam \pi^+$
(with the addition of a random $\gamma$), and a second order polynomial for the
remaining background. 
The fit gives $1597 \pm 67$ $\Lambda_{c}^+\ra \si \pi^+$ decays. 
The relative reconstruction efficiency found to be  
$\epsilon(\Lambda_{c}^+\ra \si K^+)/\epsilon(\Lambda_{c}^+\ra \si\pi^+) 
 = 0.84$ in the MC: we then calculate
$$
   \frac{\mathcal{B}(\Lambda_{c}^+\ra \si K^+)}
  {\mathcal{B}(\Lambda_{c}^+\ra \si\pi^+)} = 0.056\pm 0.014 \pm 0.008.
$$ 


\section{Observation of the {\boldmath $\Lambda_{\lowercase {c}}^+ \ra \sig K^+ \pi^-$} decay}

\begin{figure}[h]
\centering
\begin{picture}(550,200)
\put(80,40){\rotatebox{90}{\large \bf Entries/(3 MeV/{\boldmath $c^2$})}}
\put(135,0){\boldmath $M(\sig K^+ \pi^-)-M(\sig)+1.189,~~~[{\rm GeV}/c^2]$} 
\put(110,10){\includegraphics[width=0.6\textwidth]{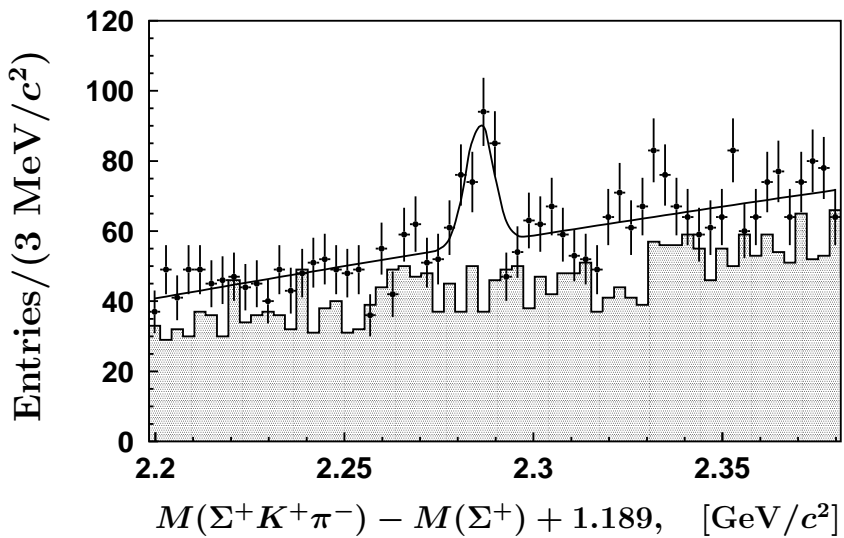}}
\end{picture}
\caption{$\Lambda_{c}^+ \ra \sig K^+ \pi^-$: invariant mass spectrum of the
	selected $\sig K^+ \pi^-$  combinations. The shaded histogram
	shows the equivalent spectrum for the $\sig$ sidebands. The mass difference
        for the sidebands is corrected using the central value of the corresponding
        sideband interval.}
\label{fig_lamc_sigkpi}
\end{figure}

The first evidence for the Cabibbo-suppressed decay $\Lambda_{c}^+ \ra \sig K^+ \pi^-$
was published by the NA32 collaboration in 1992~\cite{NA32_lamc_sigkpi}:
they found 2 events in the signal region. 
Reconstructing $\sig K^+ \pi^-$ combinations with the cuts of
Section~\ref{section-selection} tightened to require
$x_p > 0.6$, we see a clear signal peak at the $\Lambda_{c}^+$ mass,
as shown in Fig.~\ref{fig_lamc_sigkpi}.
The tighter cut is used to suppress the large combinatorial background.
We also form $\sig K^+ \pi^-$ combinations using ``$\sig$'' candidates
from mass sidebands (two 10~MeV/$c^2$ intervals centered $20$~MeV/$c^2$
below and above the nominal $\sig$ mass~\cite{PDG}), shown 
with the shaded histogram: no enhancement is seen near the $\Lambda_{c}^+$ mass.

The mass distribution is fitted with a Gaussian 
for the signal (with width fixed to $3.6$~MeV$/c^2$ from the MC) and a second
order polynomial for the background:
we find $105 \pm 24$ $\Lambda_{c}^+ \ra \sig K^+ \pi^-$ events. 
For normalization we reconstruct $\Lambda_{c}^+ \ra \sig \pi^+ \pi^-$ decays
with the same cuts, finding $2368\pm 89$ events. 
The relative efficiency of the $\Lambda_{c}^+ \ra \sig K^+ \pi^-$ channel
reconstruction with respect to $\Lambda_{c}^+ \ra \sig \pi^+ \pi^-$ 
is found to be $0.94$ in the MC. Using this value,
we obtain 
$$ 
   \frac{\mathcal{B}(\Lambda_{c}^+ \ra \sig K^+ \pi^-)}
  {\mathcal{B}(\Lambda_{c}^+ \ra \sig \pi^+ \pi^-)} = 0.047 \pm 0.011 \pm 0.008.
$$

\section{Measurement of the {\boldmath $\Lambda_{\lowercase {c}}^+ \ra \sig K^+ K^-$} and 
{\boldmath $\sig \phi$} decays}
\label{section-sigkk}

The decays $\Lambda_{c}^+ \ra \sig K^+ K^-$ and $\Lambda_{c}^+ \ra \sig \phi$
proceed dominantly via W-exchange diagrams, and were observed by CLEO in
1993~\cite{CLEO_lamc_sigkk}.
Here we measure these decay channels with improved accuracy 
and provide the first evidence
for the $\Lambda_{c}^+ \ra \Xi(1690)^0 K^+$ decay.

\begin{figure}[h]
\centering
\begin{picture}(550,200)
\put(80,40){\rotatebox{90}{\large \bf Entries/(2 MeV/{\boldmath $c^2$})}}
\put(135,0){\boldmath $M(\sig K^+ K^-)-M(\sig)+1.189,~~~[{\rm GeV}/c^2]$} 
\put(110,10){\includegraphics[width=0.6\textwidth]{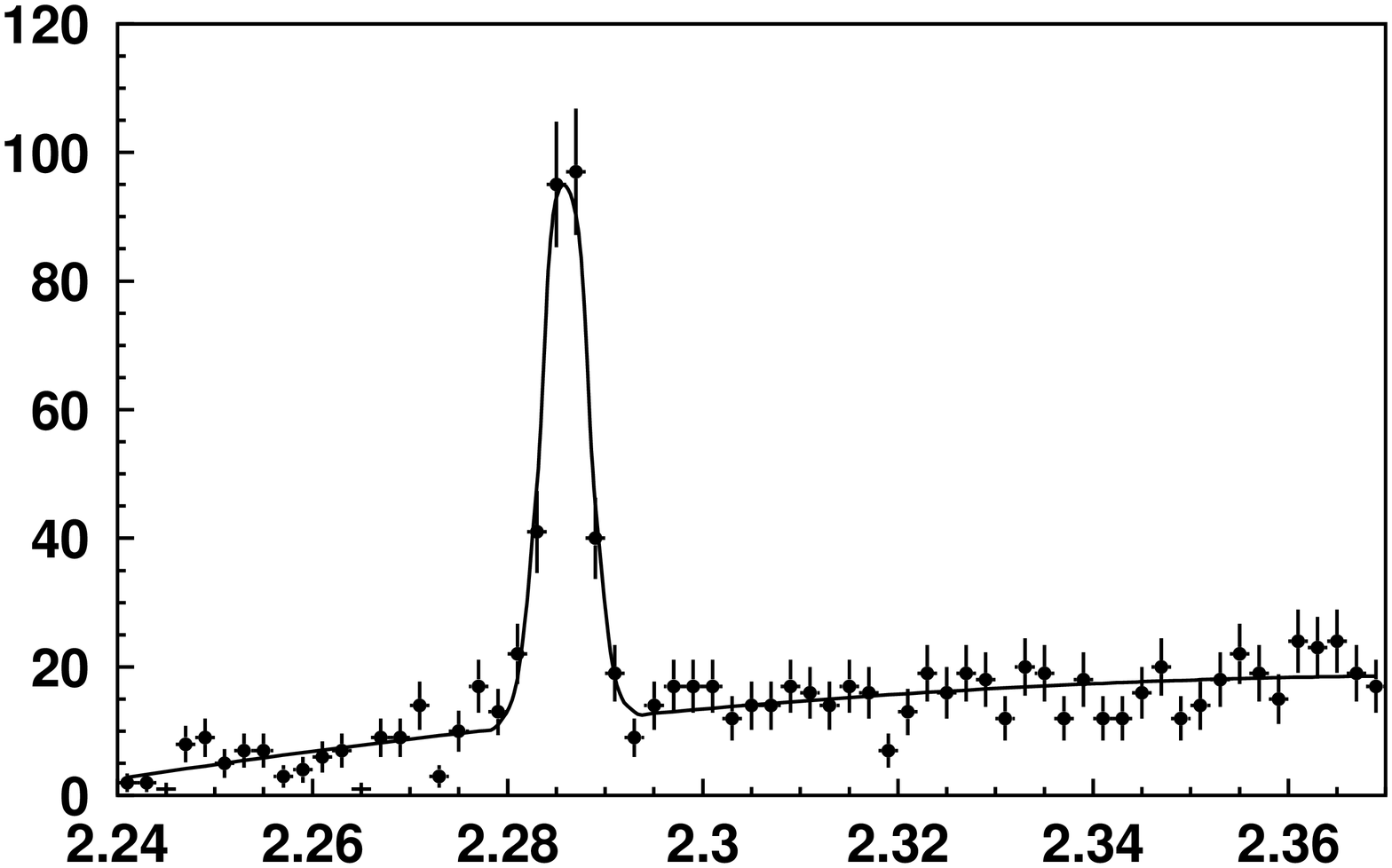}}
\end{picture}
\caption{$\Lambda_{c}^+ \ra \sig K^+ K^-$: invariant mass spectrum of the selected
	$\sig K^+ K^-$  combinations.}  
\label{fig_lamc_sigkk}
\end{figure}

Figure~\ref{fig_lamc_sigkk} shows the invariant mass spectrum for
$\Lambda_{c}^+ \ra \sig K^+ K^-$ combinations selected according to
Section~\ref{section-selection}.
A clear peak is seen at the $\Lambda_{c}^+$ mass, over a low background.
We fit the distribution using a Gaussian (with width fixed to 2.2~MeV$/c^2$
from the MC) plus a second order polynomial: the fit yields $246\pm 20$
 $\Lambda_{c}^+ \ra \sig K^+ K^-$ decays.
For normalization we reconstruct the $\Lambda_{c}^+ \ra \sig \pi^+ \pi^-$ decay
mode with equivalent cuts, 
and fit
the distribution with a Gaussian and a second order polynomial: we find 
$3650\pm 138$ $\Lambda_{c}^+ \ra \sig \pi^+ \pi^-$ events.
The relative efficiency of the $\Lambda_{c}^+ \ra \sig K^+ K^-$ decay 
reconstruction with respect to the $\Lambda_{c}^+ \ra \sig \pi^+ \pi^-$ 
decay is calculated by MC simulation and is found to be $0.89$. 
We thus extract 
$$
\frac{\mathcal{B}(\Lambda_{c}^+ \ra \sig K^+ K^-)}{\mathcal{B}(\Lambda_{c}^+ \ra \sig \pi^+ \pi^-)} = 
0.076 \pm 0.007 \pm 0.009.
$$

\begin{figure}[h]
\centering
\begin{picture}(550,200)
\put(80,40){\rotatebox{90}{\large \bf Entries/(2 MeV/{\boldmath $c^2$})}}
\put(195,0){\boldmath $M(K^+ K^-),~~~[{\rm GeV}/c^2]$} 
\put(110,10){\includegraphics[width=0.6\textwidth]{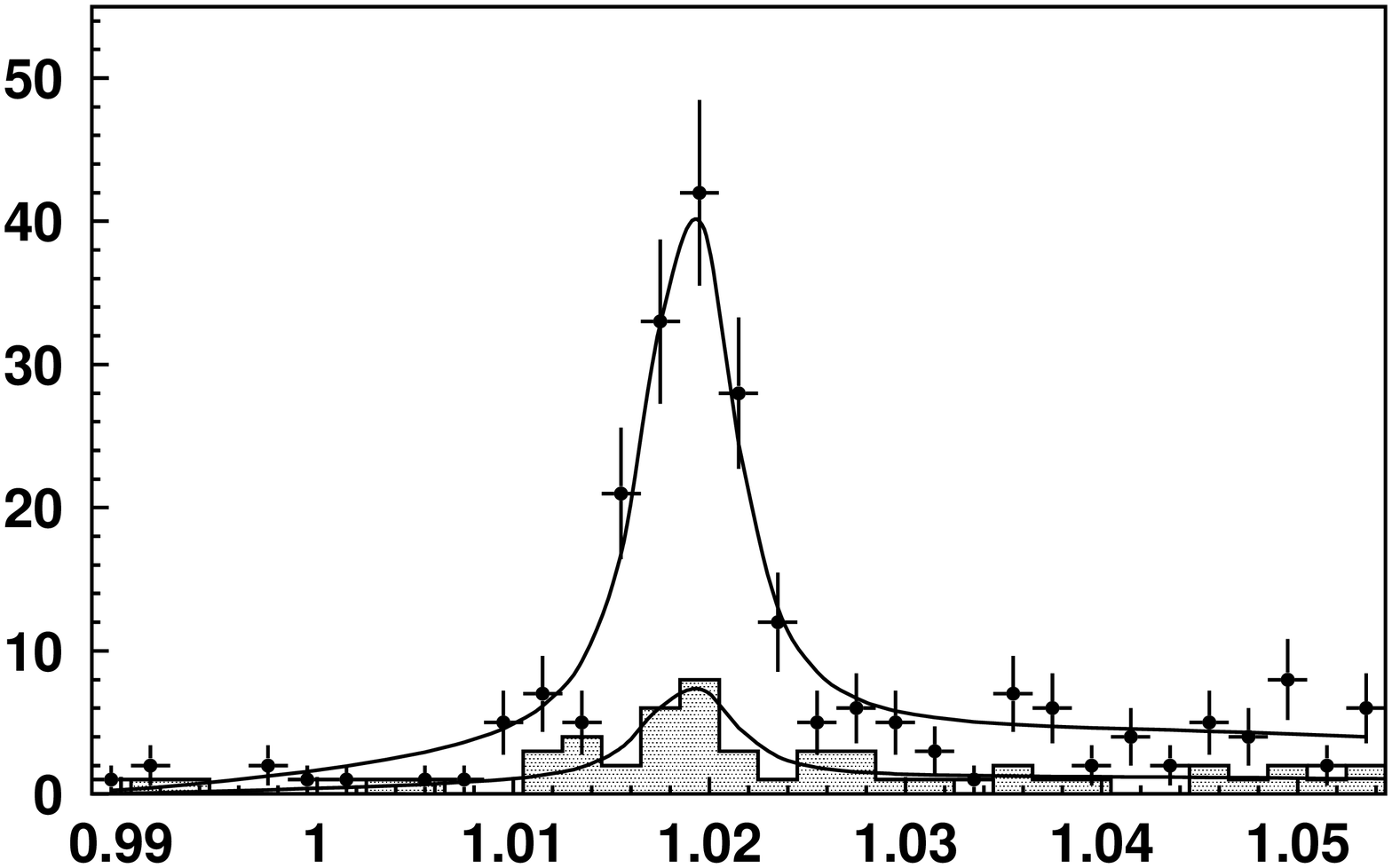}}
\end{picture}
\caption{Fitting for the $\Lambda_{c}^+ \ra \sig \phi$ component:
	the invariant mass spectra of $K^+ K^-$ combinations from
	the $\Lambda_{c}^+ \ra \sig K^+ K^-$ signal area (points with error bars)
	and $\Lambda_{c}^+$ sidebands (shaded histogram) are shown.}  
\label{fig_lamc_sigkk_mkk}
\end{figure}

In order to obtain the $\Lambda_{c}^+ \ra \sig \phi$ signal,
we take $\sig K^+ K^-$ from a $\pm 5$~MeV$/c^2$ window around the fitted
$\Lambda_{c}^+$ mass (2286~MeV$/c^2$), and plot the invariant mass of the 
$K^+ K^-$ combination, as shown in Fig.~\ref{fig_lamc_sigkk_mkk} (points with error bars);
the equivalent distribution is also shown for $\sig K^+ K^-$ in 5~MeV/$c^2$ 
sidebands
centered 12.5~MeV$/c^2$ below and above the fitted $\Lambda_{c}^+$ mass
(shaded histogram).
The distributions are fitted with a Breit-Wigner function (describing the $\phi$ signal)
convolved with a Gaussian of fixed width
(representing the detector mass resolution)
plus a second order polynomial multiplied by a square root threshold factor.
The intrinsic width of the $\phi$ Breit-Wigner function is fixed to its
nominal value~\cite{PDG}, and the width of the Gaussian resolution is
fixed to 1.0~MeV$/c^2$ based on the MC simulation.
The fit yields $153 \pm 15$ events for the $\phi$ signal in the $\Lambda_{c}^+$
region and $27 \pm 7$ in the $\Lambda_{c}^+$ sidebands. 
To extract the $\Lambda_{c}^+ \ra \sig \phi$ contribution we subtract the $\phi$
yield in the sidebands from the yield in the $\Lambda_{c}^+$ signal region,
correcting for the phase space factor obtained from the $\sig K^+ K^-$
background fitting function.
After making a further correction for the missing signal outside the
$\Lambda_{c}^+$ mass interval,
we obtain $129 \pm 17$ $\Lambda_{c}^+ \ra \sig \phi$ decays.

The relative efficiency of the $\Lambda_{c}^+ \ra \sig \phi$ reconstruction
with respect to $\Lambda_{c}^+ \ra \sig \pi^+ \pi^-$ 
is calculated using the MC and found to be $0.84$. 
Taking into account the $\phi$ branching fraction
$\mathcal{B}(\phi \ra K^+ K^-) = (49.4 \pm 0.7)\%$~\cite{PDG},
we calculate 
$$
   \frac{\mathcal{B}(\Lambda_{c}^+ \ra \sig \phi)}
  {\mathcal{B}(\Lambda_{c}^+ \ra \sig \pi^+ \pi^-)}
  = 0.085 \pm 0.012 \pm 0.012.
$$

\begin{figure}[h]
\centering
\begin{picture}(550,200)
\put(80,40){\rotatebox{90}{\large \bf Entries/(4 MeV/{\boldmath $c^2$})}}
\put(145,0){\boldmath $M(\sig K^-)-M(\sig)+1.189,~~~[{\rm GeV}/c^2]$} 
\put(110,10){\includegraphics[width=0.6\textwidth]{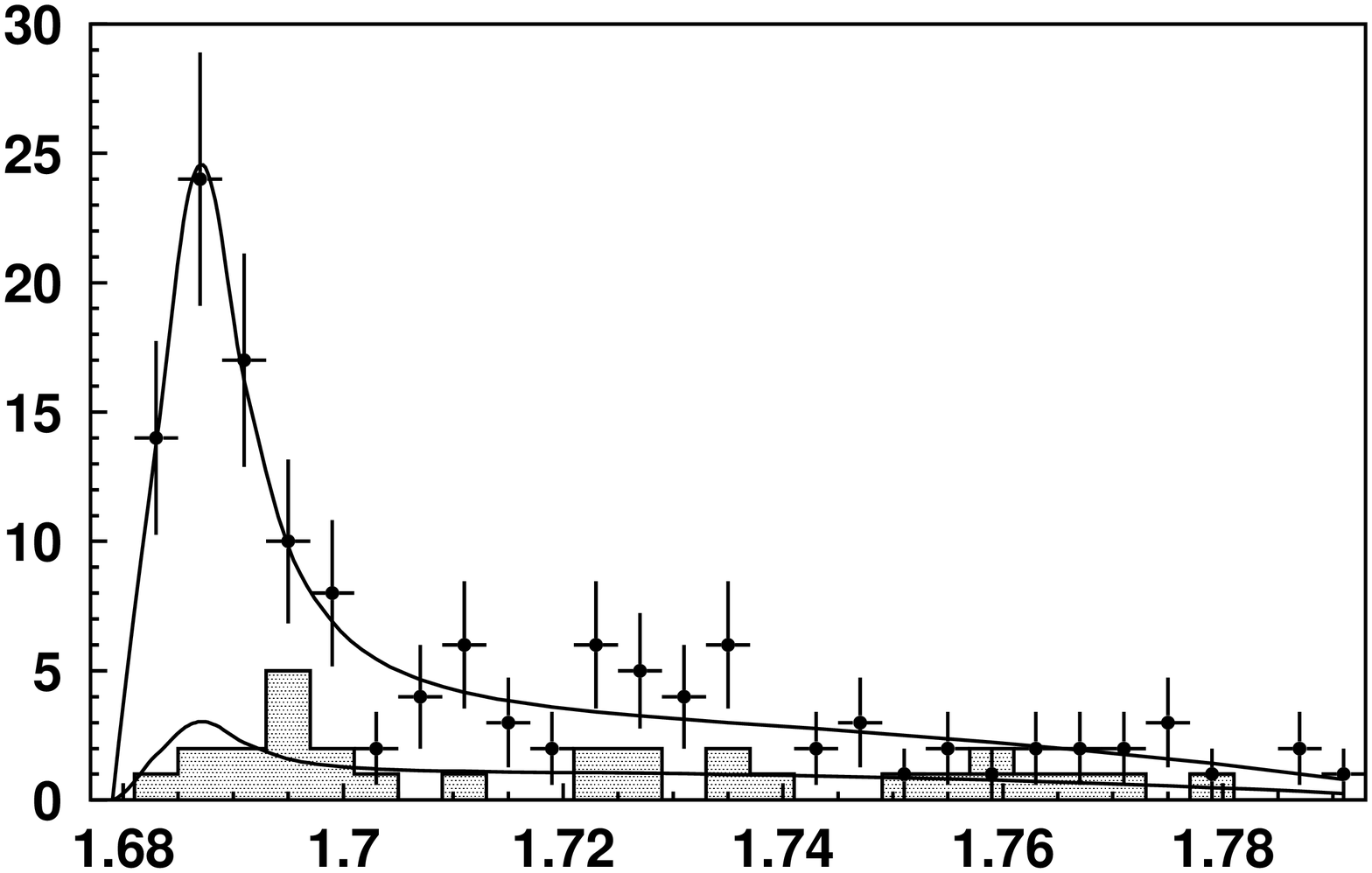}}
\end{picture}
\caption{Fitting for the $\Lambda_{c}^+ \ra \Xi(1690)^0 K^+$ component:
	the invariant mass spectrum of $\sig K^-$ combinations from the
	$\Lambda_{c}^+ \ra \sig K^+ K^-$ signal area (points with error bars) 
	and $\Lambda_{c}^+$ sidebands (shaded histogram) are shown, with
	the $\phi \ra K^+ K^-$ signal region excluded in both cases.}  
\label{fig_lamc_sigkk_sigk}
\end{figure}

We also search for resonant structure in the $\sig K^-$ system in these
decays. Figure~\ref{fig_lamc_sigkk_sigk} shows the $\sig K^-$ invariant mass spectrum for $\sig K^+ K^-$
combinations in a $\pm 5$~MeV$/c^2$ interval around the fitted $\Lambda_{c}^+$
mass (data points):
we also require $| M(K^+ K^-) - m_\phi | > 10\, \mathrm{MeV}/c^2$ to suppress
$\phi \ra K^+ K^-$.
Also shown is the $\sig K^-$ invariant mass spectrum from $\sig K^+ K^-$ combinations
selected inside $5\, \mathrm{MeV}/c^2$ sideband intervals centered 12.5~MeV$/c^2$
below and above the fitted $\Lambda_{c}^+$ mass (shaded histogram).
The $\sig K^-$ mass distribution shows evidence for the $\Xi(1690)^0$
resonant state. In order to extract this resonant contribution the histograms
are fitted with a relativistic Breit-Wigner function (describing the $\Xi(1690)^0$ signal)
plus a $(M_{\rm max} - M)^\alpha$ function multiplied by a square root threshold factor
(here $M_{\rm max}$ is the maximal allowed value of the $\sig K^-$ invariant mass).
The fit yields $82 \pm 15$ events for the $\Xi(1690)^0$ signal in the
$\Lambda_{c}^+$ region, with a fitted mass ($1688 \pm 2$)~MeV$/c^2$
and width ($11 \pm 4$)~MeV in good agreement with
previous measurements of the $\Xi(1690)^0$ parameters~\cite{PDG}.
To fit the sidebands, the function parameters are fixed to the central values
obtained from the signal fit, and both the signal 
and background normalizations are floated. A yield of $9 \pm 4$ events is found.

The $\Lambda_{c}^+ \ra \Xi(1690)^0 K^+$ contribution is obtained by subtracting
the $\Xi(1690)^0$ yield in the sidebands from the yield in the $\Lambda_{c}^+$
signal region, correcting the sideband contribution using the 
phase space factor obtained from the $\sig K^+ K^-$ background fitting function.
After a further correction for the missing signal outside the
$\Lambda_{c}^+$ mass interval, we obtain
$75 \pm 16$ $\Lambda_{c}^+ \ra \Xi(1690)^0 K^+$ decays.
We then find
$$
   \frac{\mathcal{B}(\Lambda_{c}^+ \ra \Xi(1690)^0 K^+)}
 	{\mathcal{B}(\Lambda_{c}^+  \ra \sig \pi^+ \pi^-)}
   \times
   \mathcal{B}(\Xi(1690)^0 \ra \sig K^-)
   = 0.023 \pm 0.005 \pm 0.005;
$$
the possible effects due to interference with 
$\Lambda_{c}^+ \ra \sig \phi$ are included in the systematic error
(see the discussion in Section~\ref{errors}).

Finally, the non-resonant $\Lambda_{c}^+ \ra \sig K^+ K^-$ contribution is
estimated by making invariant mass cuts
$| M(K^+ K^-) - m_\phi |	> 10\, \mathrm{MeV}/c^2$ and
$| M(\sig K^-) - M_{\Xi(1690)^0}| > 20\, \mathrm{MeV}/c^2$ to suppress the
$\phi$ and $\Xi(1690)^0$ contributions (here, $M_{\Xi(1690)^0}$ is the fitted
$\Xi(1690)^0$ mass). The resulting $\sig K^+ K^-$ mass spectrum
is 
fitted with a Gaussian (with width fixed to 2.2~MeV$/c^2$ from the MC) plus a
second order polynomial. 
The fit yields $34 \pm 9$ events.
Integrating the $\phi$ Breit-Wigner function over the allowed $M(K^+ K^-)$
region, we find that $14\%$ of the total $\Lambda_{c}^+ \ra \sig \phi$ signal
contributes to this sample: $18\pm 3$ events.
The contribution of the $\Xi(1690)^0$ mass tails is estimated to be
approximately 12$\%$ of the fitted $\Xi(1690)^0$ signal: $9 \pm 2$ events.
Subtracting these contributions, $7 \pm 10$ non-resonant events remain.
The phase space correction factor to account for the missing region around the
$\phi$ and $\Xi(1690)^0$ masses is found to be 1.63 by MC simulation of the
non-resonant $M(K^+ K^-)$  spectrum. Applying this correction we obtain
$11 \pm 16$ $\Lambda_{c}^+ \ra \sig K^+ K^-$ non-resonant decays.
Taking into account the systematic error, we obtain an upper limit 
$$
   \frac{\mathcal{B}(\Lambda_{c}^+ \ra \sig K^+ K^-)_{\mbox{\scriptsize \rm non-res}}}
  {\mathcal{B}(\Lambda_{c}^+  \ra \sig \pi^+ \pi^-)}
  < 0.018
$$
at the 90\% confidence level, 
including the possible effects due to interference with 
$\Lambda_{c}^+ \ra \sig \phi$ in the systematic errors
(see Section~\ref{errors}).


\section{Evidence for the {\boldmath $\Xi(1690)^0$} 
in
{\boldmath $\Lambda_{\lowercase {c}}^+ \ra \Lambda^0 {\bar K^0} K^+$} decays}
\label{lamkk}

\begin{figure}[h]
\centering
\begin{picture}(550,200)
\put(80,40){\rotatebox{90}{\large \bf Entries/(5 MeV/{\boldmath $c^2$})}}
\put(145,0){\boldmath $M(\lam K^0_{\rm S})-M(\lam)+1.116,~~~[{\rm GeV}/c^2]$} 
\put(110,10){\includegraphics[width=0.6\textwidth]{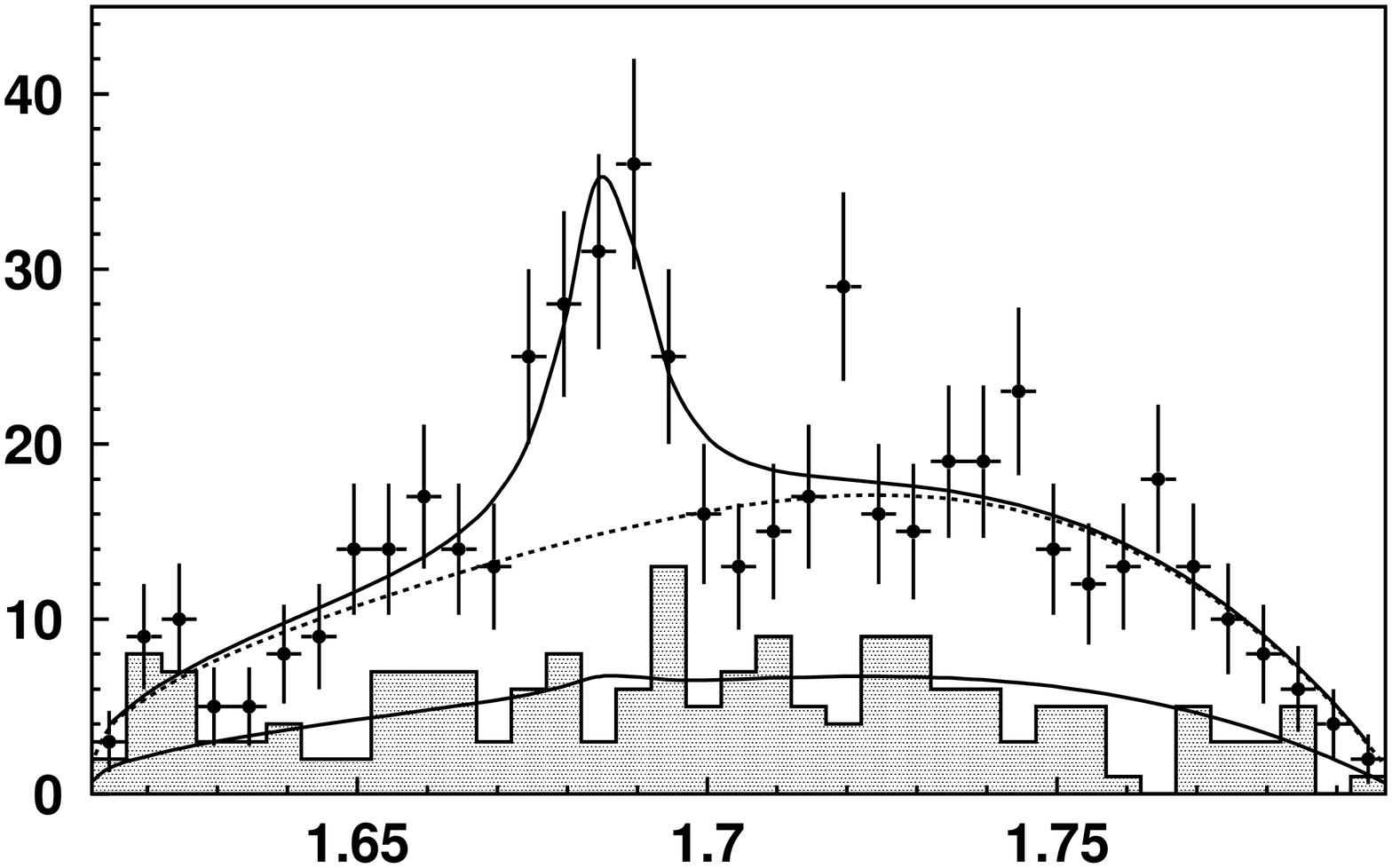}}
\end{picture}
\caption{Fitting for the $\Lambda_{c}^+ \ra \Xi(1690)^0 K^+$ component:
	the invariant mass spectrum of $\lam K^0_{\rm S}$ combinations from the
	$\Lambda_{c}^+ \ra \lam K^0_{\rm S} K^+$ signal area (points with error bars) 
	and $\Lambda_{c}^+$ sidebands (shaded histogram) are shown.
        The dashed curve represents the background function.}
\label{fig_lamc_lamk_lamkk}
\end{figure}

Another possible decay mode of the $\Xi(1690)^0$ resonant state is $\Xi(1690)^0 \ra \lam {\bar K^0}$.
Hence we have searched for the decay $\Lambda_{c}^+ \ra \Xi(1690)^0 K^+$ by reconstructing 
$\Lambda_{c}^+ \ra \Lambda^0 K^0_{\rm S} K^+$ decays and looking at the $\Lambda^0 K^0_{\rm S}$ invariant
mass distribution. Reconstructing $\Lambda^0 K^0_{\rm S} K^+$ combinations with the cuts of
Section~\ref{section-selection} we obtain an invariant mass spectrum,
which 
is fitted with a Gaussian for the signal and 
a second order polynomial for the background: we find $363 \pm 26$ 
$\Lambda_{c}^+ \ra \Lambda^0 K^0_{\rm S} K^+$ events. In order to obtain the $\Lambda_{c}^+ \ra \Xi(1690)^0 K^+$
signal, we 
take $\lam K^0_{\rm S} K^+$ from a $\pm 10$~MeV$/c^2$ window ($\approx 2.5 \sigma$) 
around the fitted $\Lambda_{c}^+$ mass
(2287~MeV$/c^2$), and plot the invariant mass of the $\Lambda^0 K^0_{\rm S}$ combination,
as shown in Fig.~\ref{fig_lamc_lamk_lamkk}
(points with error bars); the equivalent distribution is also shown for $\lam K^0_{\rm S} K^+$ from
$10$~MeV$/c^2$ sideband intervals centered $20$~MeV$/c^2$ below and above the fitted $\Lambda_{c}^+$ mass
(shaded histogram). A peak at the expected position is clearly seen.
We use a fitting procedure similar to that described in Section~\ref{section-sigkk} for the $\Lambda_{c}^+ \ra 
\sig \phi$ analysis. After subtraction of the sideband contribution and corrections we
obtain $93 \pm 26$ $\Lambda_{c}^+ \ra \Xi(1690)^0 K^+$ decays.
This confirms our observation of the $\Lambda_{c}^+ \ra \Xi(1690)^0 K^+$ decay:
significant signals are seen for both $\Xi(1690)^0 \ra \sig K^-$ and
$\Xi(1690)^0 \ra \lam {\bar K^0}$.

Using the normalization to the inclusive decay mode $\Lambda_{c}^+ \ra \lam {\bar K^0} K^+$
and the measured values for the $\mathcal{B}(\Lambda_c^+ \ra \sig \pi^+ \pi^-)$ and
$\mathcal{B}(\Lambda_c^+ \ra \lam {\bar K^0} K^+)$~\cite{PDG} we find
$$
   \frac{\mathcal{B}(\Lambda_{c}^+ \ra \Xi(1690)^0 K^+)}
 	{\mathcal{B}(\Lambda_{c}^+  \ra \lam {\bar K^0} K^+)}
   \times \mathcal{B}(\Xi(1690)^0 \ra \lam {\bar K^0})
   = 0.26 \pm 0.08 \pm 0.03.
$$
Using the value of the $\Lambda_{c}^+ \ra \Xi(1690)^0 K^+,~\Xi(1690)^0 \ra \sig K^-$ combined
branching ratio, obtained in Section~\ref{section-sigkk}, 
and the ratio of the normalization decay rates~\cite{PDG},
we find the following ratio of
$\Xi(1690)^0$ decay rates:
$$
   \frac{\mathcal{B}(\Xi(1690)^0 \ra \sig K^-)}
 	{\mathcal{B}(\Xi(1690)^0  \ra \lam {\bar K^0})} = 0.50 \pm 0.26.
$$

The corresponding ratio of the $\Xi(1690)^0$ decay rates quoted by~\cite{PDG} ($1.8\pm 0.6$
after isospin correction) is based on a single measurement reported in~\cite{xi1690}.
In order to check for possible interference effects, we studied the corresponding $\lam K^+$
invariant mass distribution and did not find any structure above a smooth background.
We have also searched for the $\Lambda_{c}^+ \ra \Xi(1690)^0 K^+$ decay in the 
$\Lambda_{c}^+ \ra (\Xi^- \pi^+) K^+$ decay mode, but did not find  any $\Xi(1690)^0$ signal in the
$\Xi^- \pi^+$ invariant mass spectrum, in agreement with the $\mathcal{B}(\Xi(1690)^0 \ra \Xi^- \pi^+)$
upper limit value from~\cite{xi1690}.


\section{Measurement of the {\boldmath $\Lambda_{\lowercase {c}}^+ \ra {\lowercase {p}} K^+ K^-$}
and {\boldmath $\Lambda_{\lowercase {c}}^+ \ra {\lowercase {p}} \phi$} decays}

The first evidence for the $\Lambda_{c}^+ \ra p \phi$ decay was reported by NA32 in 1990, who claimed
a signal of $2.8 \pm 1.9$ events~\cite{NA32_lamc_pphi}.
The decay $\Lambda_{c}^+ \ra p K^+ K^-$ was observed for the first time by E687
in 1993, who also obtained an upper limit for the branching ratio of
$\Lambda_{c}^+ \ra p \phi$~\cite{E687_lamc_pkk}.
The most recent statistically significant resonant analysis was published by
CLEO in 1996, 
who found the following branching ratios:
$\mathcal{B}(\Lambda_{c}^+ \ra p K^+ K^-)/\mathcal{B}(\Lambda_{c}^+ \ra p K^- \pi^+) = 0.039 \pm 0.009 \pm 0.007$
and $\mathcal{B}(\Lambda_{c}^+ \ra p \phi)/\mathcal{B}(\Lambda_{c}^+ \ra p K^- \pi^+) = 
0.024 \pm 0.006 \pm 0.003$~\cite{CLEO_lamc_pkk}.

\begin{figure}[h]
\centering
\begin{picture}(550,200)
\put(80,40){\rotatebox{90}{\large \bf Entries/(2 MeV/{\boldmath $c^2$})}}
\put(190,0){\boldmath $M(p K^+ K^-),~~~[{\rm GeV}/c^2]$} 
\put(110,10){\includegraphics[width=0.6\textwidth]{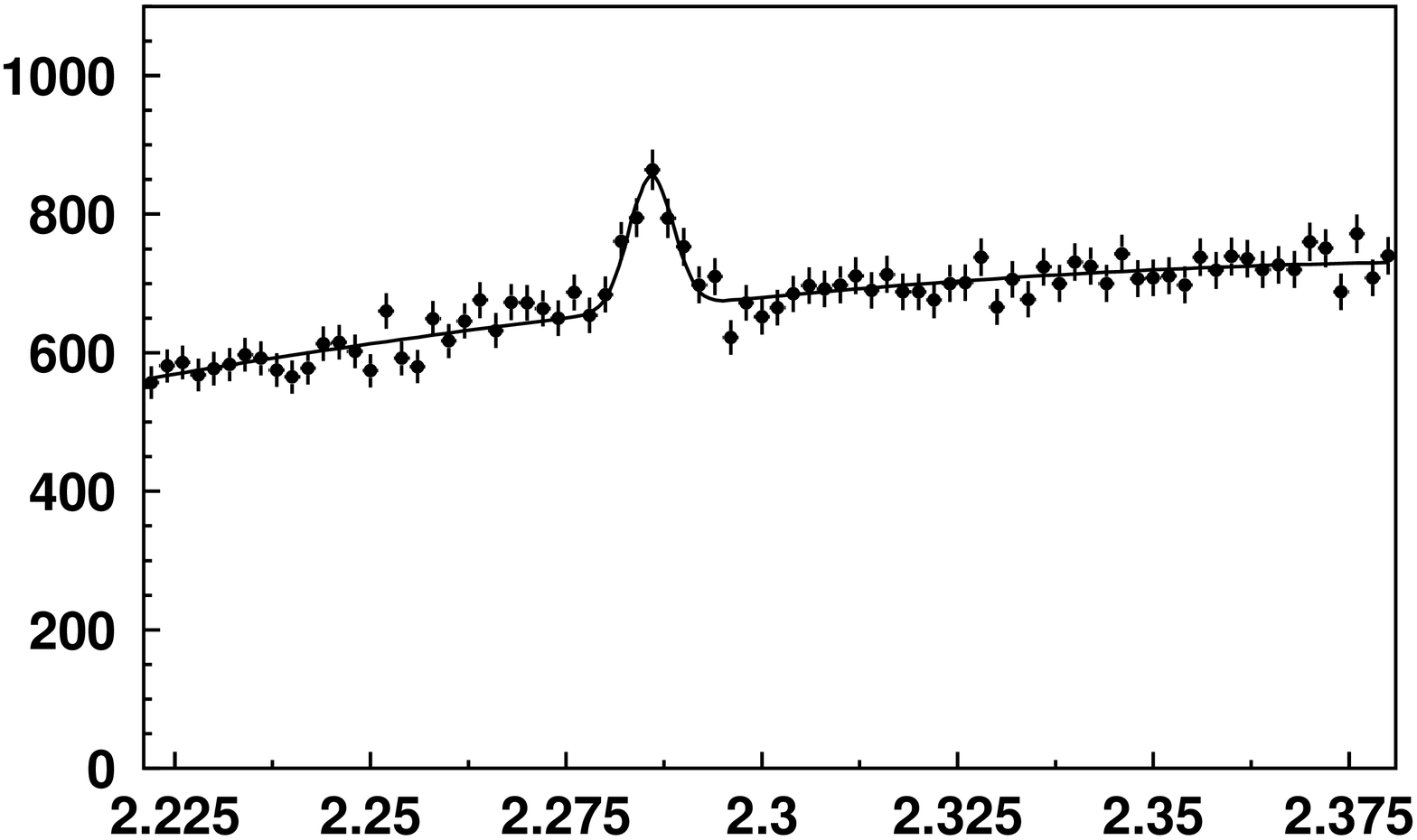}}
\end{picture}
\caption{$\Lambda_{c}^+ \ra p K^+ K^-$: invariant mass spectrum of the selected
	$p K^+ K^-$ combinations.} 
\label{fig_lamc_pkk}
\end{figure}
Reconstructing $\Lambda_{c}^+ \ra p K^+ K^-$ candidates according to
the procedure described in Section~\ref{section-selection},
we see a clear peak at the $\Lambda_{c}^+$ mass, as shown
in Fig.~\ref{fig_lamc_pkk}. 
We fit the distribution with a Gaussian
(with width fixed to $2.8\,\mathrm{MeV}/c^2$ from the MC) plus a second order
polynomial, and find $676\pm 89$ $\Lambda_{c}^+ \ra p K^+ K^-$ events.
For normalization we reconstruct the $\Lambda_{c}^+ \ra p K^- \pi^+$ decay
mode with equivalent cuts 
and fit the
distribution with a double Gaussian for the large signal peak, and a second 
order polynomial, finding $51680\pm 650$ events.
The relative efficiency of the $\Lambda_{c}^+ \ra p K^- K^+$ decay reconstruction
with respect to $\Lambda_{c}^+ \ra p K^- \pi^+$ 
is found to be $0.93$ in the MC: using this value, we extract 
$$
   \frac{\mathcal{B}(\Lambda_{c}^+ \ra p K^+ K^-)}
  {\mathcal{B}(\Lambda_{c}^+ \ra p K^- \pi^+)}
  = 0.014 \pm 0.002 \pm 0.002.
$$

\begin{figure}[h]
\centering
\begin{picture}(550,200)
\put(80,40){\rotatebox{90}{\large \bf Entries/(2 MeV/{\boldmath $c^2$})}}
\put(195,0){\boldmath $M(K^+ K^-),~~~[{\rm GeV}/c^2]$} 
\put(110,10){\includegraphics[width=0.6\textwidth]{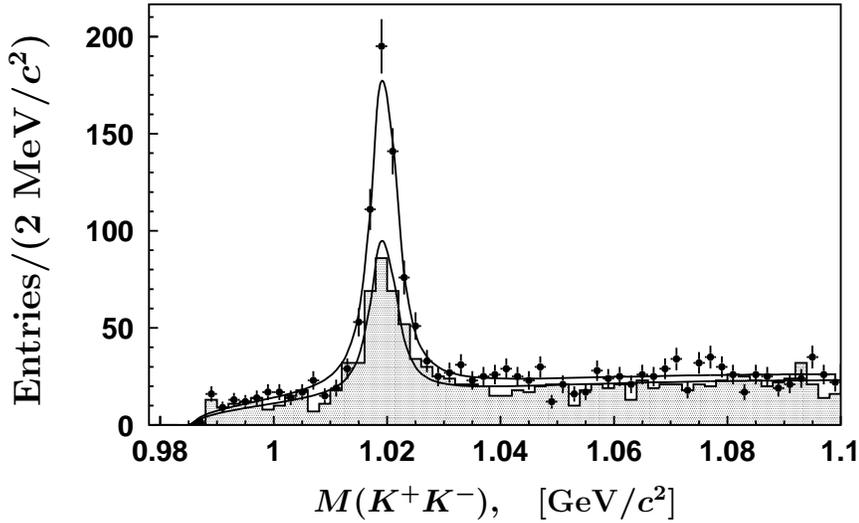}}
\end{picture}
\caption{Fitting for the $\Lambda_{c}^+ \ra p \phi$ component: 
	the invariant mass spectra of $K^+ K^-$ combinations from the
	$\Lambda_{c}^+ \ra p K^+ K^-$ signal area (points with error bars)
	and sidebands (shaded histogram).}
\label{fig_lamc_pkk_mkk}
\end{figure}

In order to obtain the $\Lambda_{c}^+ \ra p \phi$ signal we take $p K^+ K^-$
from a $\pm 6\, \mathrm{MeV}/c^2$ window
around the fitted $\Lambda_{c}^+$ mass (2286~MeV$/c^2$),
and plot the invariant mass of the $K^+K^-$ combination, as shown in
Fig.~\ref{fig_lamc_pkk_mkk} (points with error bars); 
the equivalent distribution is also shown for $p K^+ K^-$ from
$6\, \mathrm{MeV}/c^2$ sideband intervals centered $13\, \mathrm{MeV}/c^2$ below and
above the fitted $\Lambda_{c}^+$ mass (shaded histogram).
The distributions are 
fitted using a method similar to that used for the $\Lambda_{c}^+ \ra \sig \phi$ analysis
(Section~\ref{section-sigkk}).
After making a sideband subtraction and correction for the signal outside the $\Lambda_{c}^+$ mass
interval we obtain $345 \pm 43$ $\Lambda_{c}^+ \ra p \phi$ decays.

The reconstruction efficiency of the $\Lambda_{c}^+ \ra p \phi$ decay 
relative to $\Lambda_{c}^+ \ra p K^- \pi^+$ was calculated using the MC 
and found to be $0.89$. 
Using this value, we extract
$$
   \frac{\mathcal{B}(\Lambda_{c}^+ \ra p \phi)}
  {\mathcal{B}(\Lambda_{c}^+ \ra p K^- \pi^+)}
  = 0.015 \pm 0.002 \pm 0.002.
$$

The non-$\phi$ $\Lambda_{c}^+ \ra p K^+ K^-$ signal is estimated by making an
invariant mass cut $| M(K^+ K^-) - m_\phi | > 10\, \mathrm{MeV}/c^2$ to suppress
the $\phi \ra K^+ K^-$ contribution. 
%
After fitting the resulting $p K^+ K^-$ mass spectrum and applying corrections
accounting for the $\phi$ tails and the missing phase space region around the
$\phi$ mass we obtain
$344 \pm 81$ $\Lambda_{c}^+ \ra p K^+ K^-$ non-$\phi$ decays .
This corresponds to 
$$
   \frac{\mathcal{B}(\Lambda_{c}^+ \ra p K^+ K^-)_{\mbox{\scriptsize \rm non-$\phi$}}}
  {\mathcal{B}(\Lambda_{c}^+  \ra p K^- \pi^+)}
  = 0.007 \pm 0.002 \pm 0.002.
$$


\section{Systematic errors}
\label{errors}


We have considered several possible sources for the systematic errors in our measurements.
The most important is the uncertainty in the pion and kaon identification efficiencies,
which affects all ratios of signal and reference branchings. 
Based on a study of kaons and pions
from $D^{*+}$-tagged $D^0 \ra K^- \pi^+$ decays, we assign
a systematic uncertainty of $6\%$ per $K/\pi$ ratio
(\emph{e.g.}\ $6\%$ for $\lam K^+/\lam \pi^+$,  $12\%$ for $\sig K^+ K^-/\sig \pi^+ \pi^-$).

Possible biases due to fitting procedure have also been studied.
In each fit, the shape of the background function has been varied by changing
the order of the polynomial function, with any change in the signal yield being
taken as a systematic uncertainty. For each fit where the width of the signal Gaussian
was fixed to the MC prediction, we have redone the fit with a floating width,
and taken the resulting change in the yield as a systematic uncertainty.
For the $\Lambda_c^+ \ra p K^+ K^-$ and $\lam K^+$ analyses, we have assigned additional
uncertainties of $6\%$ and  $10\%$ respectively on the signal yields, based on the
fractions of signal events found in non-Gaussian tails for the normalization modes $p K^- \pi^+$
and $\lam \pi^+$ (the $p K^+ K^-$ and $\lam K^+$ samples are too small to fit for the 
presence of non-Gaussian tails).
 
For the Breit-Wigner fit of the $\phi$ signal we have varied the function by letting the
width of the convolved Gaussian float and varying the shape of the background
parameterization. We have also included the $1.4\%$ uncertainty
of $\mathcal{B}(\phi \ra K^+ K^-)$ and varied the $\phi$ nominal width within
its error~\cite{PDG}.
In the case of $\Lambda_{c}^+ \ra \sig \phi$ and $p \phi$ 
decays there is an additional source of systematic error due
to the difference in kinematics between the signal and normalization modes.
This has been estimated to be $6\%$ for $\sig \phi$ and $4\%$ for $p \phi$,
based on the difference between the MC predictions for the 
efficiency in resonant and non-resonant cases. 

In the $\Lambda_{c}^+ \ra \Xi(1690)^0 K^+,~\Xi(1690)^0 \ra \sig K^-$ resonant analysis we neglected the
possible interference between $\Xi(1690)^0$ and $\phi$ contributions. MC studies show
that this leads to an uncertainty of less than $5\%$,
due to phase space limitations in the interference region.


\section{Conclusions}

In summary, we report the first observation of
the Cabibbo-suppressed decays $\Lambda_{c}^+\ra \lam K^+$ and
$\Lambda_{c}^+\ra \si K^+$,
and the first observation of $\Lambda_{c}^+\ra \Sigma^+ K^+\pi^-$ with
large statistics.
The decays $\Lambda_{c}^+ \ra p K^+ K^-$, 
$\Lambda_{c}^+\ra p \phi$ and $\Lambda_{c}^+\ra (p K^+ K^-)_{\mbox{\scriptsize \rm non-$\phi$}}$,  
and the W-exchange decays $\Lambda_{c}^+ \ra \sig K^+ K^-$ and
$\Lambda_{c}^+ \ra \sig \phi$ have been measured with the best accuracy to date.
We have also observed evidence for the decay $\Lambda_{c}^+ \ra \Xi(1690)^0 K^+$
and set an upper limit on the
non-resonant decay mode $\Lambda_{c}^+ \ra \sig K^+ K^-$.
The results for these decay modes are listed in Table 1.

\section{Acknowledgment}

We wish to thank the KEKB accelerator group for the excellent
operation of the KEKB accelerator. We acknowledge support from the
Ministry of Education, Culture, Sports, Science, and Technology of Japan
and the Japan Society for the Promotion of Science; the Australian
Research
Council and the Australian Department of Industry, Science and
Resources; the Department of Science and Technology of India; the BK21
program of the Ministry of Education of Korea and the CHEP SRC
program of the Korea Science and Engineering Foundation; the Polish
State Committee for Scientific Research under contract No.2P03B 17017;
the Ministry of Science and Technology of Russian Federation; the
National Science Council and the Ministry of Education of Taiwan; and
the U.S. Department of Energy.



{\rotatebox{-90}{
\begin{tabular}{lcccccc}
 & & & & & & \\
$\Lambda_c^+$ signal mode & 
Signal & 
$\Lambda_c^+$ reference & 
Reference &
Relative &
$\mathcal{B}_{\rm signal}/\mathcal{B}_{\rm reference}$ & 
Other measurements  \\[-5pt]
 & 
yield & 
mode & 
yield &
efficiency &
 & 
 \\ 
\hline
$\lam K^+$ & 
$265 \pm 35$ & 
$\lam \pi^+$ &
$4550 \pm 111$ &
0.79 &
$0.074\pm 0.010\pm 0.012$ & 
$-$ \\
$\si K^+$ &
$75 \pm 18$ & 
$\si \pi^+$ &
$1597 \pm 67$ &
0.84 &
$0.056\pm 0.014\pm 0.008$ & 
$-$ \\
$ \Sigma^+ K^+ \pi^-$  & 
$105 \pm 24$ &
$ \Sigma^+ \pi^+ \pi^-$ &
$2368 \pm 89$ &
0.94 &
$0.047 \pm 0.011 \pm 0.008$ &  
$0.24^{+0.24}_{-0.16}$~\cite{NA32_lamc_sigkpi} \\
$ \Sigma^+ K^+ K^-$ &
$246 \pm 20$ & 
$ \Sigma^+ \pi^+ \pi^-$ &
$3650 \pm 138$ &
0.89 &
$0.076 \pm 0.007 \pm 0.009$ & 
$0.094 \pm 0.017 \pm 0.019$~\cite{CLEO_lamc_sigkk,CLEO_lamc_sigpipi} \\
$\Sigma^+ \phi$ & 
$129 \pm 17$ &
$ \Sigma^+ \pi^+ \pi^-$ &
$3650 \pm 138$ &
0.84 &
$0.085 \pm 0.012 \pm 0.012$ & 
$0.094 \pm 0.033 \pm 0.025$~\cite{CLEO_lamc_sigkk,CLEO_lamc_sigpipi} \\
$ \Xi(1690)^0 K^+,~\Xi(1690)^0 \ra  \sig K^-$ &
$75 \pm 16$ & 
$\Sigma^+ \pi^+ \pi^-$ &
$3650 \pm 138$ &
0.89 &
$0.023 \pm 0.005 \pm 0.005$ & 
$-$ \\
$ \Xi(1690)^0 K^+,~\Xi(1690)^0 \ra  \lam {\bar K^0}$ &
$93 \pm 26$ &
$\lam {\bar K^0} K^+$ &
$363 \pm 26$ &
1.00 & 
$0.26 \pm 0.08 \pm 0.03$ & 
$-$ \\
${ \Sigma^+ K^+ K^-}$ (\rm non-res) & 
$-$ &
$ \Sigma^+ \pi^+ \pi^-$ & 
$3650 \pm 138$ &
0.89 &
$< 0.018$ @ 90\% CL & 
$-$ \\
$ p K^+ K^-$ &
$676 \pm 89$ & 
$ p K^- \pi^+$ &
$51680 \pm 650$ &
0.93 & 
$0.014 \pm 0.002 \pm 0.002$ & 
$0.039 \pm 0.009 \pm 0.007$~\cite{CLEO_lamc_pkk} \\
$ p \phi$ & 
$345 \pm 43$ &
$ p K^- \pi^+$ &
$51680 \pm 650$ &
0.89 & 
$0.015 \pm 0.002 \pm 0.002$ & 
$0.024 \pm 0.006 \pm 0.003$~\cite{CLEO_lamc_pkk} \\
${ p K^+ K^-}$ (non-$\phi$) &
$344 \pm 81$ & 
$ p K^- \pi^+$ &
$51680 \pm 650$ &
0.93 & 
$0.007 \pm 0.002 \pm 0.002$ & 
$-$ \\
\end{tabular}
}}
{\rotatebox{-90}{
The last column shows the most accurate previous measurement
of each decay mode, where applicable.}}
{\rotatebox{-90}{
Table 1. Summary of the results obtained in this paper. }}


\begin{thebibliography}{99}

\bibitem{theory}
Y. Kohara, Nuovo Cim. {\bf A} 111 (1998) 67; \\
M. A. Ivanov et al., Phys. Rev. {\bf D} 57 (1998) 5632;\\
K. K. Sharma and R. C. Verma, Phys. Rev. {\bf D} 55 (1997) 7067; \\
L. Chau et al., Phys. Rev. {\bf D} 54 (1996) 2132;\\
A. Datta, {hep-ph/9504428};\\
T. Uppal et al., Phys. Rev. {\bf D} 49 (1994) 3417;\\ 
P. Zenczykowski, Phys. Rev. {\bf D} 50 (1994) 402;\\ 
J. K\"orner and M. Kr\"amer, Z. Phys. {\bf C} 55 (1992) 659.

\bibitem{PDG}
D. E. Groom et al.,
EPJ {\bf C} 15 (2000) 1 and 2001 off-year partial update 

\bibitem{BELLE_DETECTOR} 
A. Abashian et al. (Belle Collaboration), KEK Progress Report 2000--4 (2000),\\
to be published in Nucl. Instr. Meth. {\bf A}. 

\bibitem{NA32_lamc_sigkpi}
S. Barlag et al. (NA32 Collaboration),
Phys. Lett. {\bf B} 283 (1992) 465. 

\bibitem{CLEO_lamc_sigkk}
P. Avery et al. (CLEO Collaboration),
Phys. Rev. Lett. 71 (1993) 2391. 

\bibitem{xi1690}
C. Dionisi et al.,
Phys. Lett. {\bf B} 80 (1978) 145. 

\bibitem{NA32_lamc_pphi}
S. Barlag et al. (NA32 Collaboration),
Z. Phys. {\bf C} 48 (1990) 29. 

\bibitem{E687_lamc_pkk}
P. L. Frabetti et al. (E687 Collaboration),
Phys. Lett. {\bf B} 314 (1993) 477. 

\bibitem{CLEO_lamc_pkk}
J. P. Alexander et al. (CLEO Collaboration),
Phys. Rev. {\bf D} 53 (1996) R1013. 

\bibitem{CLEO_lamc_sigpipi}
Y. Kubota et al. (CLEO Collaboration),
Phys. Rev. Lett. 71 (1993) 3255. 

\end{thebibliography}
\end{document}